\documentclass[fleqn,usenatbib]{mnras}

\usepackage{textgreek}
\usepackage{flushend}

\usepackage{newtxtext,newtxmath}

\usepackage[T1]{fontenc}

\DeclareRobustCommand{\VAN}[3]{#2}
\let\VANthebibliography\thebibliography
\def\thebibliography{\DeclareRobustCommand{\VAN}[3]{##3}\VANthebibliography}

\usepackage{graphicx}	
\usepackage{amsmath}	

\title[Intensity Interferometer with Linear-mode APDs]{{\scshape Sirius}: A Prototype Astronomical Intensity Interferometer Using Avalanche Photodiodes in Linear Mode}

\author[J. Oh et al.]{
Junghwan Oh,$^{1,2}$
Jan Wagner,$^{2,3}$
Sascha Trippe,$^{1}$\thanks{E-mail: trippe@astro.snu.ac.kr}
Taeseok Lee,$^{1}$
Bangwon Lee$^{1,2}$
and Chang Hee Kim$^{2}$
\\
$^{1}$Seoul National University, Department of Physics and Astronomy,  1 Gwanak-ro, Gwanak-gu, Seoul 08826, Korea\\
$^{2}$Korea Astronomy and Space Science Institute, 776 Daedeok-daero, Yuseong-gu, Daejeon 34055, Korea\\
$^{3}$Max-Planck-Institut f\"ur Radioastronomie, Auf dem H\"ugel 69, 53121 Bonn, Germany
}

\date{Accepted XXX. Received YYY; in original form ZZZ}

\pubyear{2020}

\begin{document}
\label{firstpage}
\pagerange{\pageref{firstpage}--\pageref{lastpage}}
\maketitle

\begin{abstract}
Optical intensity interferometry, developed in the 1950s, is a simple and inexpensive method for achieving angular resolutions on microarcsecond scales. Its low sensitivity has limited intensity interferometric observations to bright stars so far. Substantial improvements are possible by using avalanche photodiodes (APDs) as light detectors. Several recent experiments used APDs in single-photon detection mode; however, these either provide low electronic bandwidths (few MHz) or require very narrow optical bandpasses. We present here the results of laboratory measurements with a prototype astronomical intensity interferometer using two APDs observing an artificial star in continuous (``linear'') detection mode with an electronic bandwidth of 100~MHz. We find a photon--photon correlation of about $10^{-6}$, as expected from the ratio of the coherence times of the light source and the detectors. In a configuration where both detectors are on the optical axis (zero baseline), we achieve a signal-to-noise ratio of $\sim$2700 after 10 minutes of integration. When measuring the correlation as a function of baseline, we find a Gaussian correlation profile with a standard deviation corresponding to an angular half-width of the artificial star of $0.55''$, in agreement with the estimate by the manufacturer. Our results demonstrate the possibility to construct large astronomical intensity interferometers using linear-mode APDs.
\end{abstract}

\begin{keywords}
instrumentation -- interferometers
\end{keywords}


\section{Introduction \label{sec:intro}}

Achieving ever higher angular resolution has been one of the main drivers for the development of ever larger optical telescopes. However, the relation $\theta \approx \lambda / D$ between resolution angle $\theta$, wavelength $\lambda$, and telescope aperture $D$ dictates that even future 30-meter class telescopes will achieve resolutions of a few milliarcseconds at best, and only when compensating atmospheric turbulence with adaptive optics systems.

Angular resolutions on sub-milliarcsecond scales require the use of interferometry. Present-day astronomical interferometers are \emph{amplitude interferometers}: light collected by two (or more) telescopes is forwarded -- via systems of mirrors, light tunnels, and/or optical fibers -- to a beam combiner. There, the light rays are superimposed coherently and the resulting interference pattern is analyzed. Mathematically, an amplitude interferometer measures the complex spatial first-order coherence function $\gamma(u,v)$, with Fourier plane coordinates $u$ and $v$, of light. The distribution of the values of $\gamma(u,v)$ in the $uv$ plane corresponds to the Fourier transform of the on-sky (with coordinates $x$, $y$) light distribution $L(x,y)$ of an astronomical source (van Cittert--Zernicke theorem). Unfortunately, astronomical amplitude interferometry is extremely challenging. Even for a small optical bandwidth $\Delta\nu = 10^{13}$~Hz (corresponding to $\Delta\lambda = 10$\,nm at $\lambda = 550$\,nm), the corresponding coherence length is $w_c \approx c/\Delta\nu \approx 30\,\mu$m (with $c$, $\nu$, and $\lambda$ being the speed of light, the frequency, and the wavelength of the light, respectively). Accordingly, errors in the optical path lengths need to be controlled down to the level of micrometers, for interferometers that span tens to hundreds of meters in size. The required level of complexity is the main reason why only a handful of astronomical optical interferometers, with maximum baselines of a few hundred meters, exist. This is in sharp contrast to radio interferometry where coherence lengths are given by the comparably narrow electronic bandwidths of radio receivers (a few GHz at most) which makes possible interferometry over intercontinental distances, with resolution angles as small as about 25~$\mu$as \citep{eht2019}. Optical interferometry has advanced our knowledge of stellar physics \citep[e.g.,][]{boyajian2012a, boyajian2012b, boyajian2013} and, very recently, of the accretion flow into the supermassive black hole in the center of the Milky Way, Sagittarius~A*, \citep{gravity2018a} and the dynamics of the broad-line region of the quasar 3C~273 \citep{gravity2018b}.

\begin{figure*}
\includegraphics[width=0.50\textwidth]{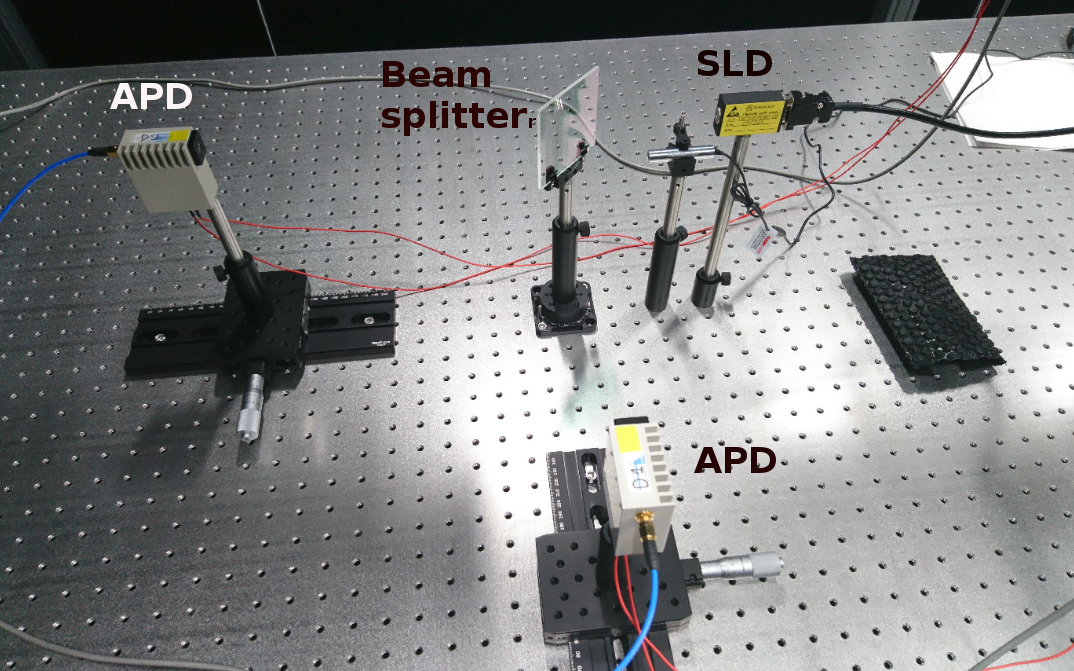}
\includegraphics[width=0.49\textwidth]{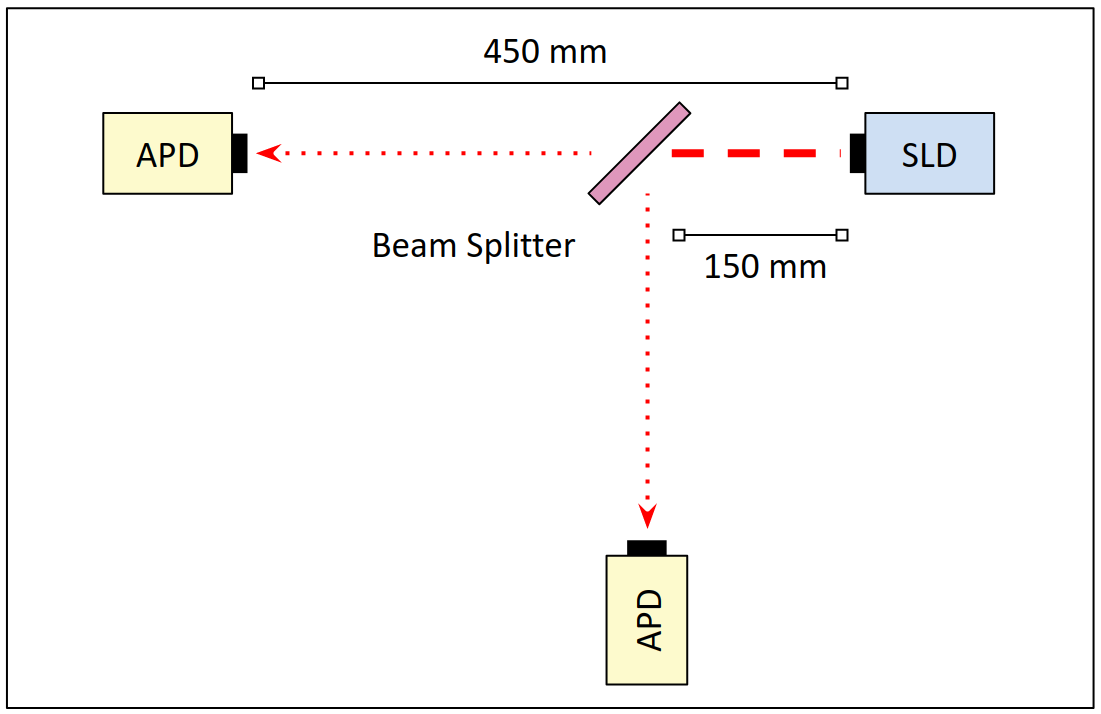} \\
\includegraphics[width=0.50\textwidth]{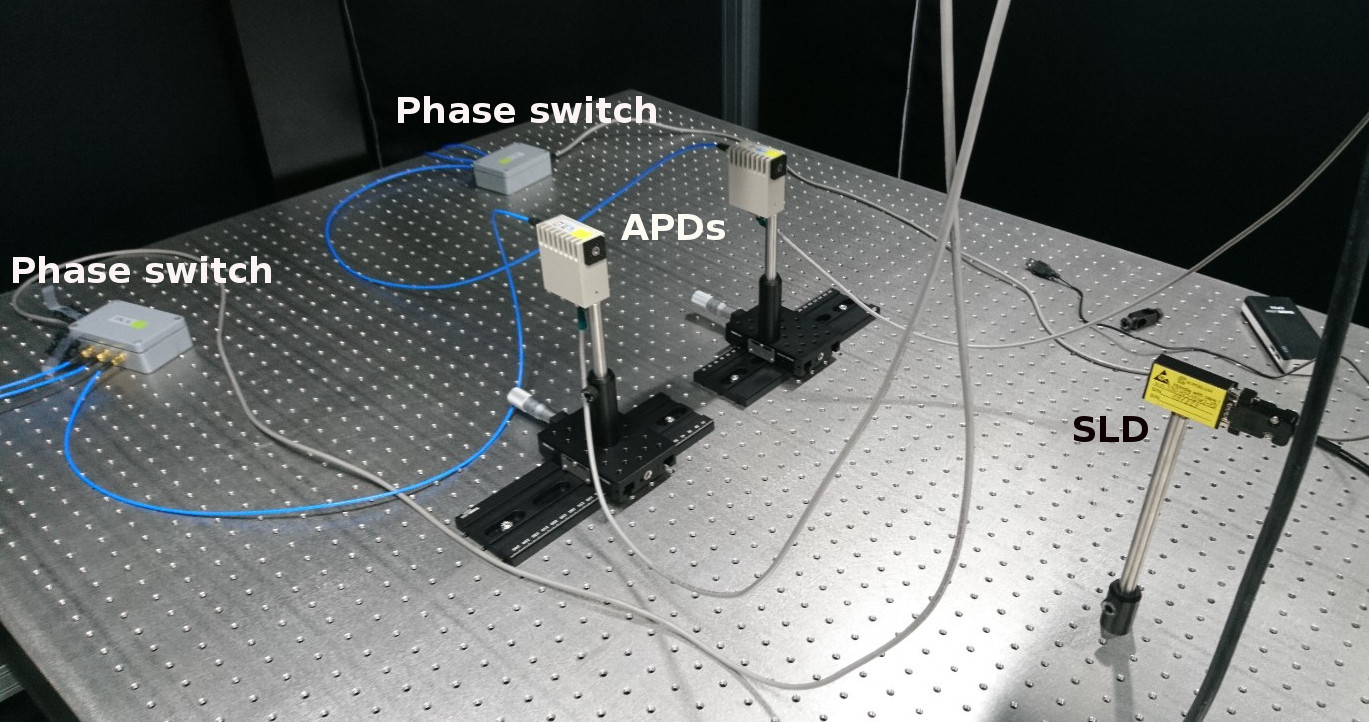}
\includegraphics[width=0.495\textwidth]{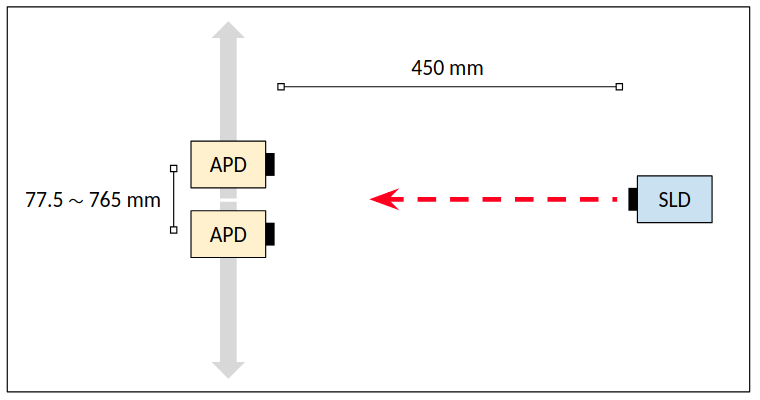}
\caption{The two interferometer configurations. \emph{Top panels:} Zero-baseline configuration. The light from the SLD is divided by a beam splitter. The APDs are both placed on the optical axis and observe the SLD with zero relative displacement. \emph{Bottom panels:} Variable-baseline configuration. The two APDs are placed on a baseline perpendicular to the optical axis and observe the SLD. The baseline can be adjusted by moving one or both APDs. Screw holes on the optical table are separated by 25\,mm.
\label{fig:configs}}
\end{figure*}

Historically, the first astronomical optical interferometer (the Narrabri Stellar Intensity Interferometer [NSII]; \citealt{hanbury1964, hanbury1967a}) made use of \emph{intensity interferometry}. Intensity interferometry is based on the Hanbury Brown--Twiss effect \citep{hanbury1957a, hanbury1957b, twiss1959}: when correlating the fluctuations in the intensities of light measured at two different locations (i.e., correlating the intensities after subtracting their time averages), $\Delta I_1, \Delta I_2$, one finds the squared absolute value of the first-order coherence function,
\begin{equation}
|\gamma(u,v)|^2 = \frac{\langle\Delta I_1\Delta I_2\rangle}{\langle I_1\rangle\langle I_2\rangle}
\label{eq:coherence}
\end{equation}
where $I_1, I_2$ are the intensities measured at two locations and $\langle...\rangle$ denotes time averaging. In contrast to amplitude interferometry, it is not possible to derive the target image $L(x,y)$ from $|\gamma(u,v)|^2$ via the van Cittert--Zernicke theorem because the phase information is lost. The target image can, at least in principle, be reconstructed either from triple correlations of intensities \citep{gamo1963, sato1978, wentz2015} or by application of the Cauchy--Riemann equations to the $|\gamma(u,v)|^2$ distribution \citep{holmes2004, nunez2012a, nunez2012b}.

The advantage of intensity interferometry over amplitude interferometry is its technical simplicity. In an intensity interferometer, light is collected by two (or more) telescopes. At each telescope, the light is detected by a photo detector with a high (tens to hundreds of MHz) sampling rate. The output voltage from the detector is transmitted to a correlator where its fluctuations are correlated with the signal from another telescope. Like in the case of radio interferometry, but unlike the case of optical amplitude interferometry, the coherence time of the signal is given by the electronic bandwidth of the photo detectors; for a bandwidth of 100~MHz (as was the case for the NSII), the corresponding coherence length is $w_c\approx3$~m. Accordingly, intensity interferometers are highly robust against inaccuracies in the signal path; this makes baselines of arbitrary length possible. This statement also applies to the telescopes: the interferometer does not require the formation of images, only the efficient collection of photons. The telescopes therefore can be coarse light collectors (``light buckets''). Furthermore, intensity interferometry is insensitive to atmospheric turbulence \citep{tan2016}. Sampling rates of tens of MHz exceed the highest frequencies of atmospheric fluctuations at visible wavelengths by at least three orders of magnitude.

\begin{figure*}
\includegraphics[width=\textwidth]{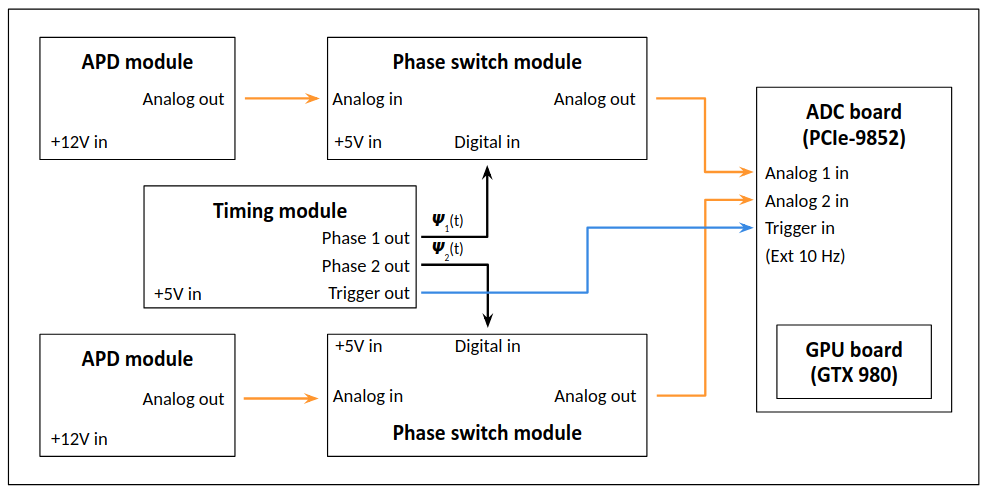}
\caption{Layout and signal processing chain of the \textsc{Sirius} interferometer. Each APD module receives the light from the artificial star and outputs an analog AC signal. This signal is passed through a phase switch with switching function $\psi(t)$ as function of time $t$. The timing module provides two orthogonal phase switching functions $\psi_{1,2}(t)$ for the two phase switches and a trigger signal for the digitizer. The analog-to-digital conversion (ADC) board accepts the signals from the two interferometer arms at two analog inputs and digitizes them. The digitized signals are correlated by the software correlator, the calculations are run on the GPU board.
\label{fig:layout}}
\end{figure*}

The principal disadvantage of intensity interferometry is its low sensitivity. The response time of even the fastest photo detectors is much longer than the coherence time of the light. Using, again, the example of an optical bandpass of $\Delta\lambda=10$\,nm at $\lambda=550$\,nm and an electronic bandpass of 100\,MHz, the frequency ratio of the two bandpasses, and thus (inversely) of the coherence times, is $10^5$. This implies that the intensity interferometer reduces the coherence of the light by a similar factor. For unpolarized light, the signal-to-noise ratio of an intensity interferometer following from photon statistics (in SI units) is given by
\begin{equation}
\left(\frac{S}{N}\right)_{\rm s} = \alpha\,A\,n_{\nu}\,\eta\,\Gamma^2(u,v)\,\left(\frac{t\,\Delta f}{2}\right)^{1/2}
\label{eq_snii}
\end{equation}
where $\alpha\in[0,1]$ is the quantum efficiency of the detector; $A$ is the collecting area of one light collector; $n_{\nu}$ is the number of photons per time, per unit area, and per unit optical bandwidth; $\eta\in[0,1]$ is the efficiency of the instrument; $t$ is the integration time; and $\Delta f$ is the electronic bandwidth in Hz \citep[c.f.][Section~3.4]{trippe2014}. The normalized correlation factor $\Gamma^2$ is defined \citep[cf. Equation (3.1) of][]{hanbury1958iii} like
\begin{equation}
\Gamma^2(u,v) = \frac{C(u,v)}{C(0,0)}
\label{eq:Gamma}
\end{equation}
where $C(u,v)$ is the degree of correlation measured at $uv$ plane coordinates $(u,v)$. Assuming that realistic observations require $S/N\gtrsim5$ for $\Gamma^2(u,v)=1$ and observing times $t=1$\,h, the limiting photometric magnitude for an intensity interferometer with the technical parameters of the NSII is $m_X\approx2.5$, $X=V,R,I$. For this reason, the NSII was shut down in 1972 after completing measurements of the angular diameters of 32 stars. Since then, development efforts have focused on optical amplitude interferometry.

As one can read off Equation~(\ref{eq_snii}), substantially increasing the performance of an intensity interferometer beyond the one of the NSII requires (i) larger collecting areas, and/or (ii) higher instrumental efficiencies, especially a higher quantum efficiency and wider electronic bandwidth. In the following, we focus on option (ii). Limited to technology of the 1970s, the NSII used photomultiplier tubes (PMTs) with $\alpha\approx20$\% and $\Delta f \approx 100$\,MHz. Much better values -- up to $\alpha\approx85$\% around $\lambda=700$\,nm and $\Delta f \approx 1$\,GHz -- have been reached by silicon avalanche photodiodes (APDs; \citealt{renker2007}). All else equal, these values are sufficient to boost the sensitivity of an NSII-type interferometer by a factor of about 13, corresponding to 2.8 photometric magnitudes. Furthermore, APDs have diameters of few millimeters, whereas the apertures of PMTs measure several centimeters. Their small size makes it possible to group APDs into linear arrays. When placed inside a spectrograph, each pixel (i.e., each individual APD) can be illuminated with light of a different color. Since the sensitivity of an intensity interferometer depends on the number of photons \emph{per unit optical bandwidth}, the signal-to-noise ratio of the instrument increases in proportion to the square root of the number of independently correlated spectral channels; this observation gave birth to the concept of multi-channel intensity interferometry \citep{trippe2014}.

The possibility of a revival of optical intensity interferometry in astronomy has been noted by several groups. \citet{trippe2014} noted the opportunities offered by the use of avalanche photodiodes. \citet{dravins2015} observed pinholes illuminated by scattered laser light in the laboratory and demonstrated that the image of a source can be reconstructed from $|\gamma(u,v)|^2$ if the $uv$ plane is densely sampled. \citet{tan2016} measured the \emph{temporal} coherence function (intensity autocorrelation) $|\gamma(u=0,v=0,\tau)|^2$, with $\tau$ being a time delay, of sunlight. \citet{guerin2017} measured the intensity autocorrelations in the light from three stars. \citet{matthews2018} and \citet{zmija2020} calculated intensity correlations from data recorded at each detector and correlated off-line at a later time, thus demonstrating the feasibility of very long baseline intensity interferometry. \citet{guerin2018} measured the spatial intensity correlation for (a different set of) three stars by combining the signals from two 1-m telescopes separated by 15\,m.

The aforementioned experiments all used photomultiplier tubes, single-photon avalanche photodiodes (SPADs), or photon-counting setups. SPADs are APDs operated with reverse voltages higher than their breakdown voltages. This leads to a large number of photoelectrons and a high gain (up to $\sim10^6$) but introduces a dead time during which the photodiode is unresponsive; this reduces their electronic bandwidths to a few MHz typically. Photon counting requires the number of photons per time arriving at the detector (photon incidence rate) to be limited to the count rate of the detector; this makes it necessary to limit the optical bandwidth to $\lesssim0.1$~nm (e.g., \citealt{tan2016}).

To overcome those limitations, we designed a novel laboratory optical intensity interferometer dubbed \textsc{Sirius}.\footnote{Sirius A was the first star ever resolved by an intensity interferometer \citep{hanbury1958}.} Our experiment employs APDs in \emph{continuous} mode where each photon triggers an electron avalanche of tens to hundreds of photoelectrons. This reduced gain weakens the output signal and makes signal processing challenging, but ensures that the photodiode remains responsive continuously, resulting in bandwidths of hundreds of MHz -- which is the bandwidth range required for an astronomical intensity interferometer. The design of, and results from, our laboratory interferometer are discussed in the following.

\section{The Instrument \label{sec:instrument}}

\subsection{Interferometer Layout \label{sec:layout}}

Our laboratory intensity interferometer was designed as a two-element long baseline interferometer, meaning the two interferometer arms are physically disconnected except for cables for signal transmission and power supply. Our setup was located on an optical table. Two avalanche photodiodes were pointed at a light source (the ``star simulator''). The signals from the detectors were passed through phase switches and are correlated in a software correlator. We employed two different optical configurations (see Figure~\ref{fig:configs}):
\begin{enumerate}
\item A ``zero-baseline configuration'' where the light from the source is split by a beam splitter in the ratio 50:50. One detector was placed onto the optical axis along the transmitted beam, the other was placed on the optical axis of the reflected beam.
\item A ``variable-baseline configuration'' where both detectors point at the light source without a beam splitter. A specific baseline was selected by displacing the APDs perpendicular to the optical axis. The minimum baseline (77.5\,mm) was determined by the physical size of the detector modules.
\end{enumerate}
We selected and/or developed the different components such that the interferometer design was applicable to astronomical observations. The interferometer was located in an optical laboratory in the Center for Astrophysics and Space Science of Seoul National University. The individual components (see Figures \ref{fig:configs} and \ref{fig:layout}) are described in the following sub-sections.

\begin{figure}
\centering
\includegraphics[width=\columnwidth]{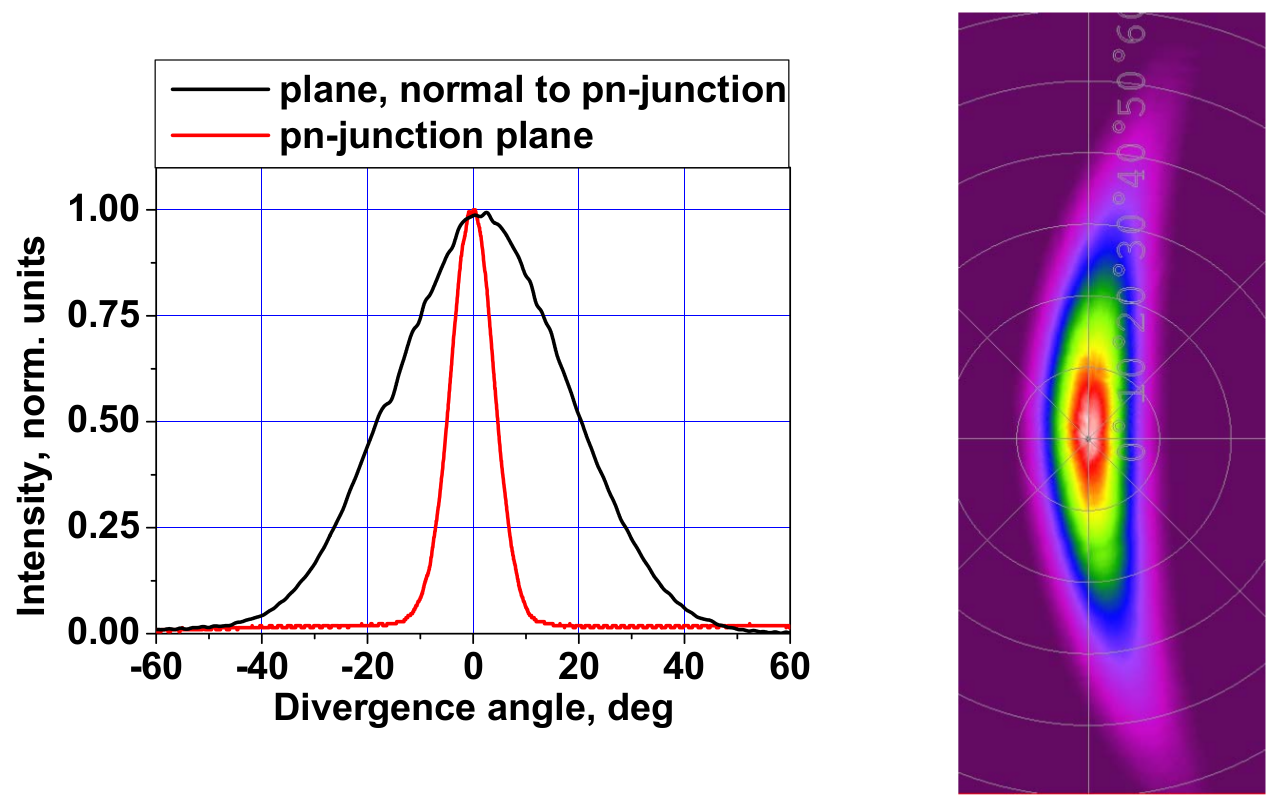}
\caption{Beam pattern of the superluminescent diode. \emph{Left:} Cross-sections through the beam parallel and perpendicular to the P-N junction plane. \emph{Right:} Two-dimensional polar map of the beam, showing its ``crescent'' shape. (Diagrams provided by Superlum) \label{fig:sldbeam}}
\end{figure}

\subsection{Star Simulator \label{sec:star}}

As indicated by Equation~(\ref{eq_snii}), achieving a high signal-to-noise ratio requires the use of a light source with high spectral flux, i.e, a high number of photons per time, per unit area, and per unit optical bandwidth. In addition, photon statistics dictates that the light source has to be incoherent; coherent light sources (lasers, masers) have $|\gamma(u,v)|^2=0$ at all $(u,v)$ positions. Previous experiments employed either high-power gas discharge lamps \citep[e.g.,][]{hanbury1957b} or randomized (scattered) laser light \citep{dravins2015}. For our experiments, we selected a superluminescent diode (SLD) from Superlum (Ireland), model model SLD-330-HP3-TOW2-PD, which provides a maximum output power of 50\,mW. After testing various parameter configurations, we eventually selected an output power of 8\,mW for our experiments. The SLD emits linearly polarized (with effective degree of polarization $m_{\rm L} \approx 85$\%) light into a platykurtic spectral window centered at about 785\,nm with a full width at zero intensity (FWZI) of about 70\,nm and a full width at half maximum (FWHM) of about 50\,nm. The active area of the diode is a few micrometers in size in each direction; measuring the actual effective size of the source was part of our experiment. The SLD emits a crescent shaped Gaussian beam with a full width at half maximum (FWHM) of about $40^{\circ}$ and $10^{\circ}$ perpendicular and parallel to the plane of the P-N junction, respectively (see Figure~\ref{fig:sldbeam}).

\subsection{Photo Detectors \label{sec:apd}}

\begin{figure}
\centering
\includegraphics[width=0.9\columnwidth]{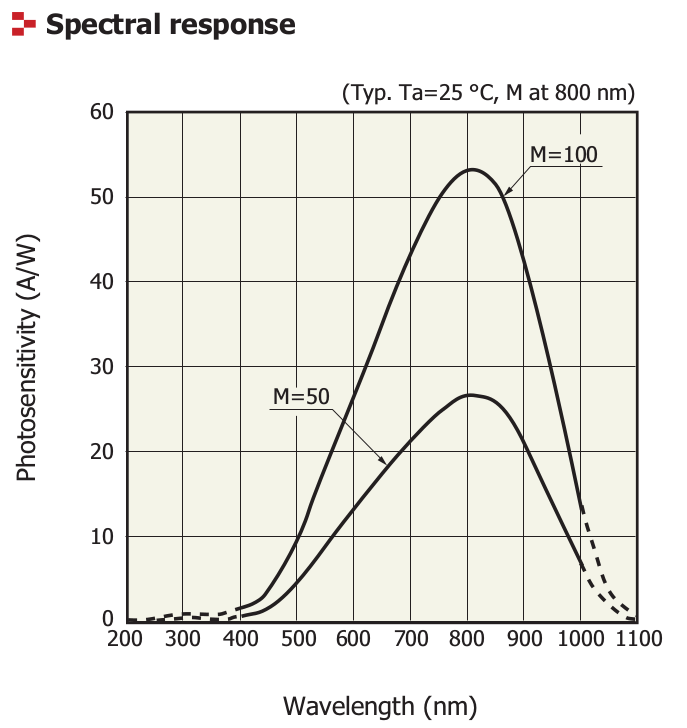} \\
\includegraphics[width=0.9\columnwidth]{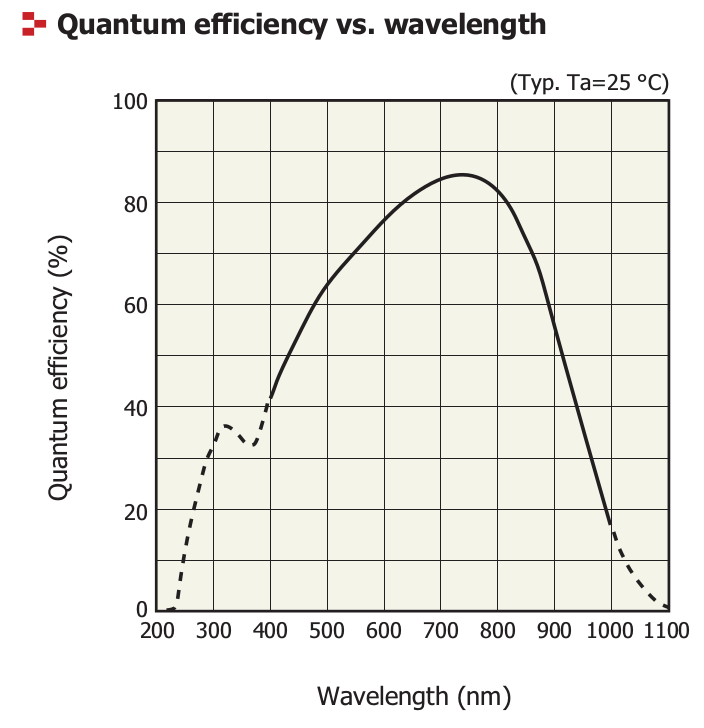}
\caption{Performance of Hamamatsu Photonics Silicon APDs as function of wavelength. \emph{Top:} Photosensitivity for two different gains $M$ of 50 and 100 (at 800\,nm); our detectors were factory-set to $M=60$. The photosensitivity peaks at about 800\,nm. \emph{Bottom:} Quantum efficiency. The maximum efficiency of about 85\% is reached at a wavelength of about 750\,nm. (Diagrams by Hamamatsu Photonics) \label{fig:apddata}}
\end{figure}

We selected two avalanche photodiodes from Hamamatsu Photonics (Japan), specifically the model S2384 in custom-made modules. The gain was factory-set to 60. Each APD was circular with an active-area diameter of 3\,mm. The photoelectric sensitivity (photocurrent per irradiated power) was approximately\,30 A/W at 800\,nm, the quantum efficiency at $\lambda=785$\,nm (i.e., the central wavelength of the light emitted by the SLD) was about 82\% (see Figure~\ref{fig:apddata}). The electronic bandwidth was 120\,MHz with the bandpass ranging from 50\,kHz to about 120\,MHz. The APD modules worked as high-pass filters which removed the direct current (DC) component of the output signals and passed on an alternating current (AC) signal that corresponds to the intensity \emph{fluctuations} of the recorded light. Since the intensity fluctuations are dominated by Poisson noise, the output voltage from the detectors is proportional to the square root of the light intensity.

\subsection{Phase Switches \label{sec: switch}}

To eliminate the influence of correlated instrumental noise, we implemented a phase switching hardware system. It was designed to decohere (average out) unwanted correlated signals affecting analog sections after APD light detection.

Our phase switching system was composed of a timing module and two phase switching modules, one for each interferometer arm (cf. Figure~\ref{fig:layout}). The timing module provided three synchronous reference signals: a trigger signal for the digitizer board with a frequency of 10\,Hz, and two phase switching signals, $\psi_{1}(t)$ and $\psi_{2}(t)$, having orthogonal periodic patterns that varied with time $t$ and repeated at 10\,Hz. After comparing the performance of various signal patterns (square function vs. Walsh functions) and frequencies, we settled for square switching functions with frequencies of 160\,Hz and 320\,Hz for interferometer arms no. 1 and 2, respectively. Phase switching modules resided next to the APDs, and passed through the analog APD signals either non-inverted (in-phase), or inverted (phase shifted by $180^{\circ}$), as controlled by the phase switching signals. Data acquisition was triggered periodically by the timing module. Before correlating the digitized signals in the software correlator the signals were digitally demodulated, with the trigger signal serving as the reference for fully synchronous (de)modulation.

\begin{figure*}
\centering
\includegraphics[width=\textwidth]{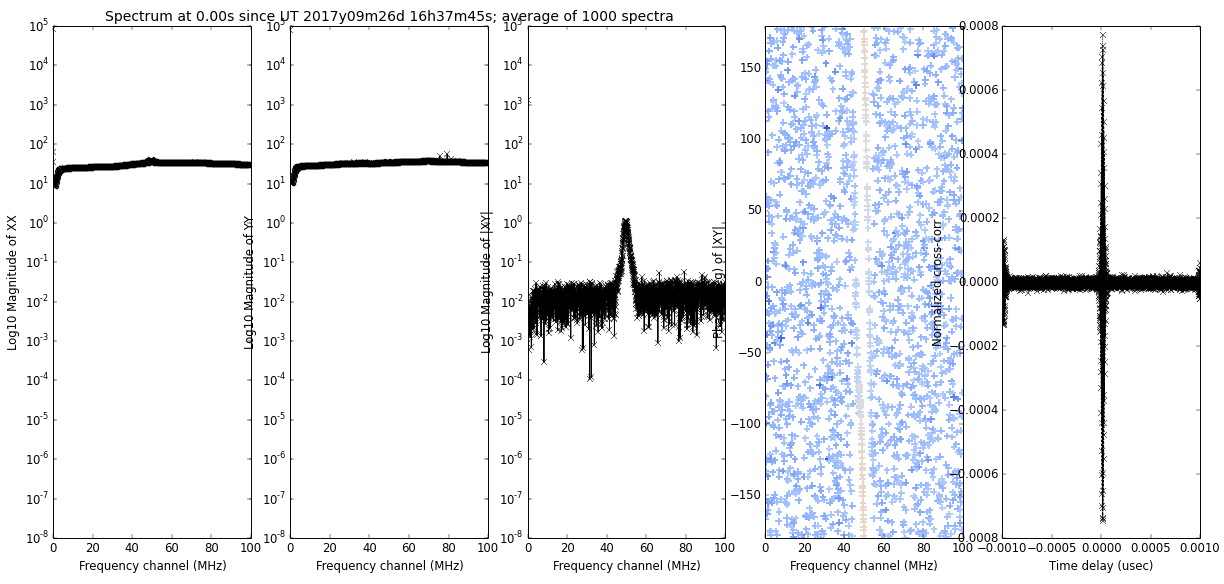}
\caption{Cross correlation spectrum analysis for the \textsc{Sirius} interferometer. Panels show from left to right: auto-correlation spectrum of the signal from detector 1; auto-correlation spectrum of the signal from detector 2; cross-correlation amplitude spectrum; cross-correlation phase spectrum; normalized cross-correlation as a function of time delay $\tau$. This example data set was taken in September 2017 during the first round of testing of the software correlator; it showed a weak but significant spurious cross-correlation signal centered at 50\,MHz. \label{fig:spectrum}}
\end{figure*}

\subsection{Digitizer \label{sec:digitizer}}


For processing in the software correlator, the phase-switched wide-band APD signals needed to digitized. We used the ADLink PCIe-9852 dual-channel 14-bit 100\,MHz analog-to-digital converter (ADC) board with a programmable input voltage range. The sampling frequency of 200\,MS/s was derived from the on-board synthesizer of the ADC board; having a common frequency reference ensured that both signals were sampled coherently and did not need corrections for relative frequency offset nor frequency drift. The ADC board was configured to capture an 80\,ms block of samples upon each external trigger event generated by the timing module (100\,ms trigger interval; 10\,Hz). Input signals were digitized at 16\,384-level 14-bit quantization. For our experiment, we programmed the analog inputs to their minimum range of $\pm0.2$\,V, corresponding to a resolution of 24.4\,\textmugreek{}V per level. The signals from the APDs had amplitudes on the order of several mV, meaning they were distributed over a few hundred levels typically. Such a number of levels ensures a quantization efficiency virtually indistinguishable from unity (cf. Table~8.2 of \citealt{thompson2017}).


For perfectly synchronous phase (de)modulation and sampling we had designed our timing module to output an ADC reference clock. Sampler boards featuring external reference clock input as well as an analog bandwidth of $2\times\ge$100\,MHz were, however, not available at the time of our measurements. The time bases of the timing module and the ADC board were thus running independently. Nevertheless, they proved sufficiently accurate and stable; the signal of a single APD split into both interferometer arms and phase modulated yielded correlation coefficients in the software correlator consistently greater than 98\%.

\subsection{Software Correlator \label{sec:correlator}}

For correlating and processing the data, we implemented a dedicated real time software correlator. The correlator captured the data in from the digitizer board and processed them on a graphics processing unit (GPU) provided by an NVIDIA GTX~980 graphics card. Digitizer, software correlator, and graphics card were all located in the same data server. To permit performance and quality checks, each 80 ms data segment was correlated using two methods.
\begin{enumerate}
\item Direct correlation. The cross correlation $\langle V_1(t) \cdot V_2(t + \tau) \rangle$ was calculated at a single relative time delay, $\tau$, by directly multiplying time domain data from the two inputs, time averaging, and normalizing. This provided the normalized cross-correlation at zero relative time delay ($\tau=0$).
\item Fourier transform (FX) correlation. The cross correlation was formed via discrete Fourier transforms according to the cross-correlation theorem (i.e., $\langle V_1(t) \cdot V_2(t + \tau) \rangle \equiv \mathcal{F}^{-1}(\tau) \{ \mathcal{F}\{V_1(t)\} \cdot \mathcal{F}\{V_2(t)\}^* \}$, where $\mathcal{F}$ denotes the continuous Fourier transform). This approach provides a cross-power spectrum, as well as the cross correlation as a function of $\tau$ (see Figure~\ref{fig:spectrum}).
\end{enumerate}
Whereas our physical signal is encoded in the normalized correlation at $\tau=0$, $C(u,v,0)$, the correlation spectra were necessary for understanding the characteristics of the correlator and to test for the presence of spurious signals. The normalized correlation (at $\tau=0$) was given by
\begin{equation}
C(u,v) = \frac{ \langle V_1(t) \cdot V_2(t) \rangle }{ \sigma_1\cdot\sigma_2 }
\label{eq:correl}
\end{equation}
where $V_{1}(t), V_{2}(t)$ were the voltages of the two AC signals as function of time $t$, $\sigma_{1}, \sigma_{2}$ were the standard deviations of $V_{1,2}(t)$, and $\langle...\rangle$ denotes averaging in time. Each 80-ms data set provided a value of $C(u,v)$ once per 100\,ms.

The second-order coherence function $|\gamma|^2$ (Equation \ref{eq:coherence}) follows from a renormalization of $C(u,v)$. The function $|\gamma(u,v)|^2$ is normalized by division by the time averaged intensities $\langle I_{1}\rangle, \langle I_{2}\rangle$; each of these intensities is proportional to the average number of photons received per detector integration time, $\langle N_{\rm ph}\rangle$. The profile function $C(u,v)$ is normalized by division by the standard deviations of the voltages, each of which is proportional to the square root of the average number of photons received per detector integration time (Poisson noise), i.e., $\sqrt{\langle N_{\rm ph}\rangle}$. The symmetry of our setup ensures that both detectors receive the same number of photons per time in average; thus, we do not need to distinguish $\langle N_{\rm ph}\rangle_i$ for detectors $i=1,2$. Eventually, it follows that
\begin{equation}
\label{eq:scaling}    
C(u,v) = |\gamma(u,v)|^2\cdot\langle N_{\rm ph} \rangle .
\end{equation}

\section{Results \label{sec:results}}

\subsection{Zero-Baseline Correlation \label{sec:zerobase}}

In the zero-baseline configuration using the beamsplitter to superimpose
the two APDs, we tested for the presence (or absence) of significant correlation signals under varying conditions. Integration times for each measurement were 10 minutes. In case of ``dark noise'' measurements, we kept both APDs in the dark. In case of ``white noise'' measurements, we illuminated both detectors with incandescent room light. In the case of ``on source'' measurements, we illuminated the APDs only with the light from the SLD. Only in case of on-source measurements, we expect to observe a physical correlation signal. In case of the dark/white-noise measurements, no correlation should be observed -- ideally.

\begin{figure}
\centering
\includegraphics[width=\columnwidth]{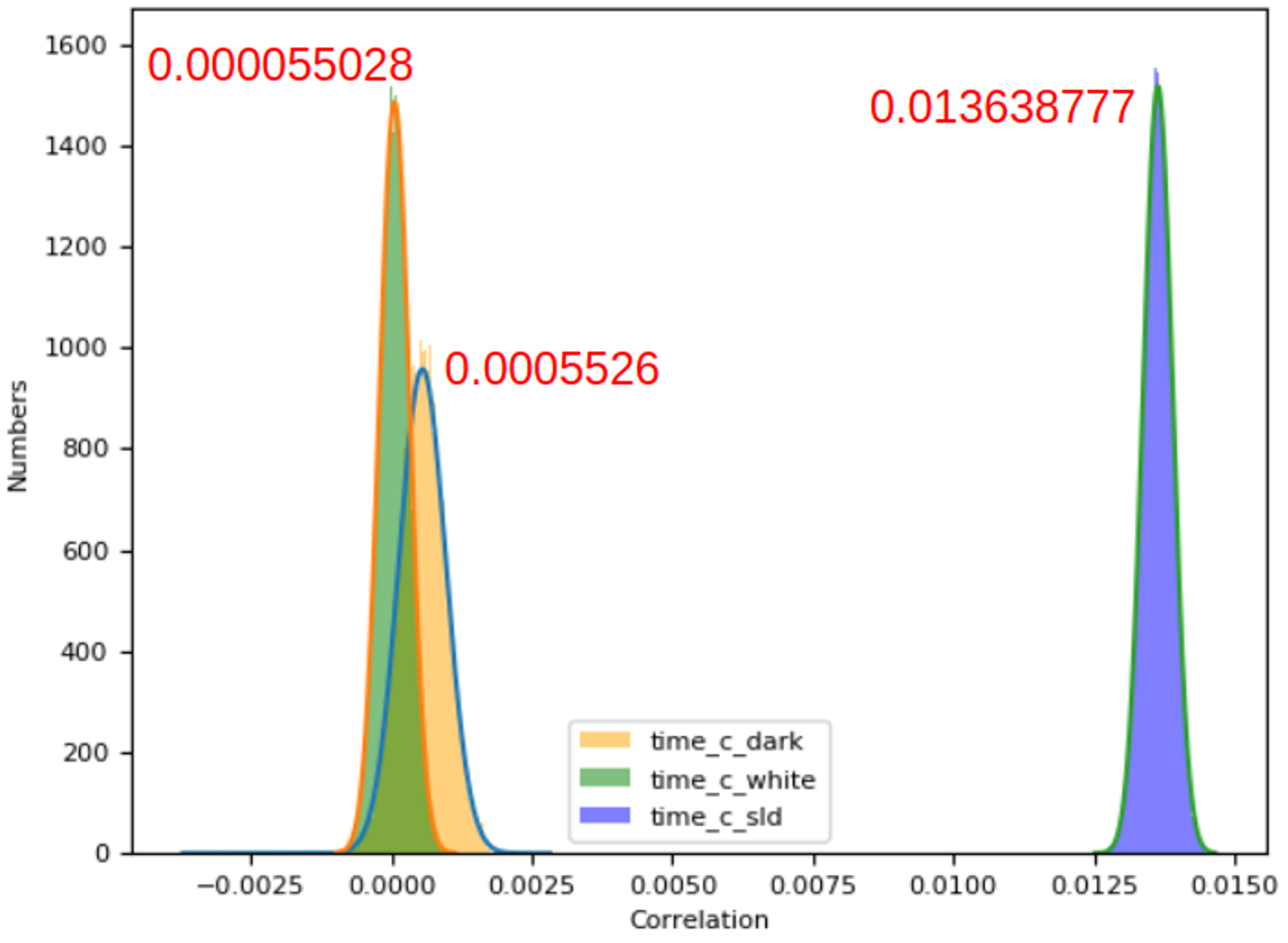}
\caption{Histograms of normalized correlations from three different test measurements in the zero-baseline configuration. Colored areas correspond to the data, continuous lines mark the best-fit Gaussian profiles. A number next to a histogram denotes its mean value. Measurements \texttt{time\_c\_white} (leftmost histogram) and \texttt{time\_c\_dark} are results obtained when the detectors were illuminated with incandescent room light and were kept dark, respectively. The measurement \texttt{time\_c\_sld} (rightmost histogram) results from observation of the SLD.  \label{fig:result-zerobase}}
\end{figure}

Figure~\ref{fig:result-zerobase} summarizes the results of our zero-baseline measurements. All measurements find correlation values $C(0,0)$ that follow Gaussian distributions. The dark/white-noise measurements find correlation values that are, in average, very close to (but significantly larger than) zero; these results show that a small bias is present in our interferometer. The mean correlation values are about $5.5\times10^{-4}$ and $5.5\times10^{-5}$ for dark and white noise, respectively. In contrast, the on-source measurement finds an average value of $C(0,0) = 0.0136$. With standard errors of mean being about $5\times10^{-6}$ for each measurement, the on-source value is about $2700\sigma$ different from zero and at least 25 times higher than the highest instrumental bias level.

Using Equation~(\ref{eq:scaling}), we convert the correlation $C(0,0)$ into the second-order coherence $|\gamma(0,0)|^2$. The light from the SLD is emitted into an elliptical beam with FWHM opening angles of $40^{\circ}\times10^{\circ}$ (see Section~\ref{sec:star}). We checked numerically that a photodetector measuring 3~mm in diameter placed on the optical axis at a distance of 450~mm receives a fraction of about $2.4\times10^{-4}$ of the flux emitted by the SLD. For a SLD power of 8~mW and with the light shared equally among two APDs by the beam splitter, each detector receives a power of about $0.96\,\mu$W, corresponding to about $3.8\times10^{12}$ photons per seconds at a wavelength of 785~nm. Since the APDs sample the light $2.4\times10^8$ times per second, the average number of photons per detector integration time is $\langle N_{\rm ph} \rangle \approx\ $16,000. Accordingly, we find $|\gamma(0,0)|^2 \approx 8.5\times10^{-7}$ for the zero-baseline correlation.

\subsection{Correlation as a Function of Baseline \label{sec:corrprofile}}

An actual astronomical interferometer probes the correlation as a function of $uv$ coordinates. In our case, we used our two-element interferometer to measure the normalized correlation as a function of baseline $b$ (which is related to the interferometric $uv$ radius like $b/\lambda = \sqrt{u^2 + v^2}$), $C(b)$. After demonstrating the successful measurement of photon--photon correlation in the zero-baseline configuration (Section~\ref{sec:zerobase}), we re-arranged our interferometer into the variable-baseline configuration. Starting from the minimum baseline given by the physical size of the detectors and their mounts, 77.5\,mm, we increased the baseline to 85\,mm  and, from then on, in steps of 10\,mm out to a maximum of 765\,mm. For each baseline we measured the normalized correlation $C(b)$ for 10 minutes. The correlation value was $C(b)=0.307$ at $b=77.5$\,mm and dropped smoothly to $C(b)=0.000167$ at $b=765$\,mm.

Taking into account the angular SLD beam profile (Section~\ref{sec:star} and Figure~\ref{fig:sldbeam}), we calculated the second-order coherence function as function of baseline,
\begin{equation}
|\gamma(b)^2| = \frac{C(b)}{2\,\langle N_{\rm ph} \rangle\,G(\phi)} ~~~~~ {\rm with} ~~~~~ G(\phi) = \exp\left[-\frac{\phi^2}{2\sigma_{\phi}^2}\right]
\label{eq:renorm}
\end{equation}
where $\langle N_{\rm ph} \rangle =16$,000, $\phi = \arctan[b/(2d)]$, $\sigma_{\phi} = 17^{\circ}$, and $d = 450$\,mm. In the variable-baseline configuration, the light is not divided by a beam splitter; therefore, each detector would receive $2\langle N_{\rm ph} \rangle$ (rather than $\langle N_{\rm ph} \rangle$) photons per detector integration time if it were placed on the optical axis.

\begin{figure}
\begin{center}
\includegraphics[width=\columnwidth]{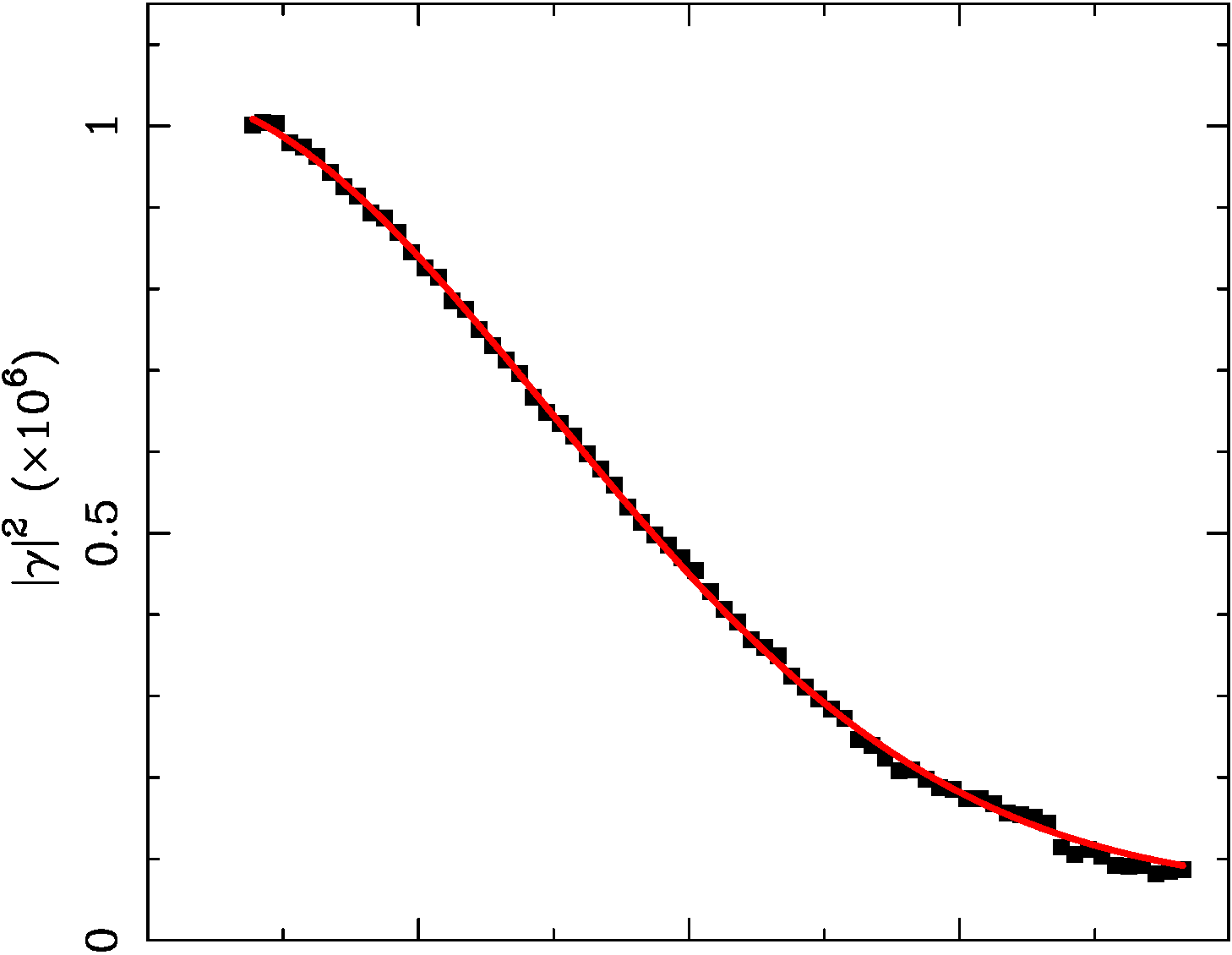} \\ \includegraphics[width=\columnwidth]{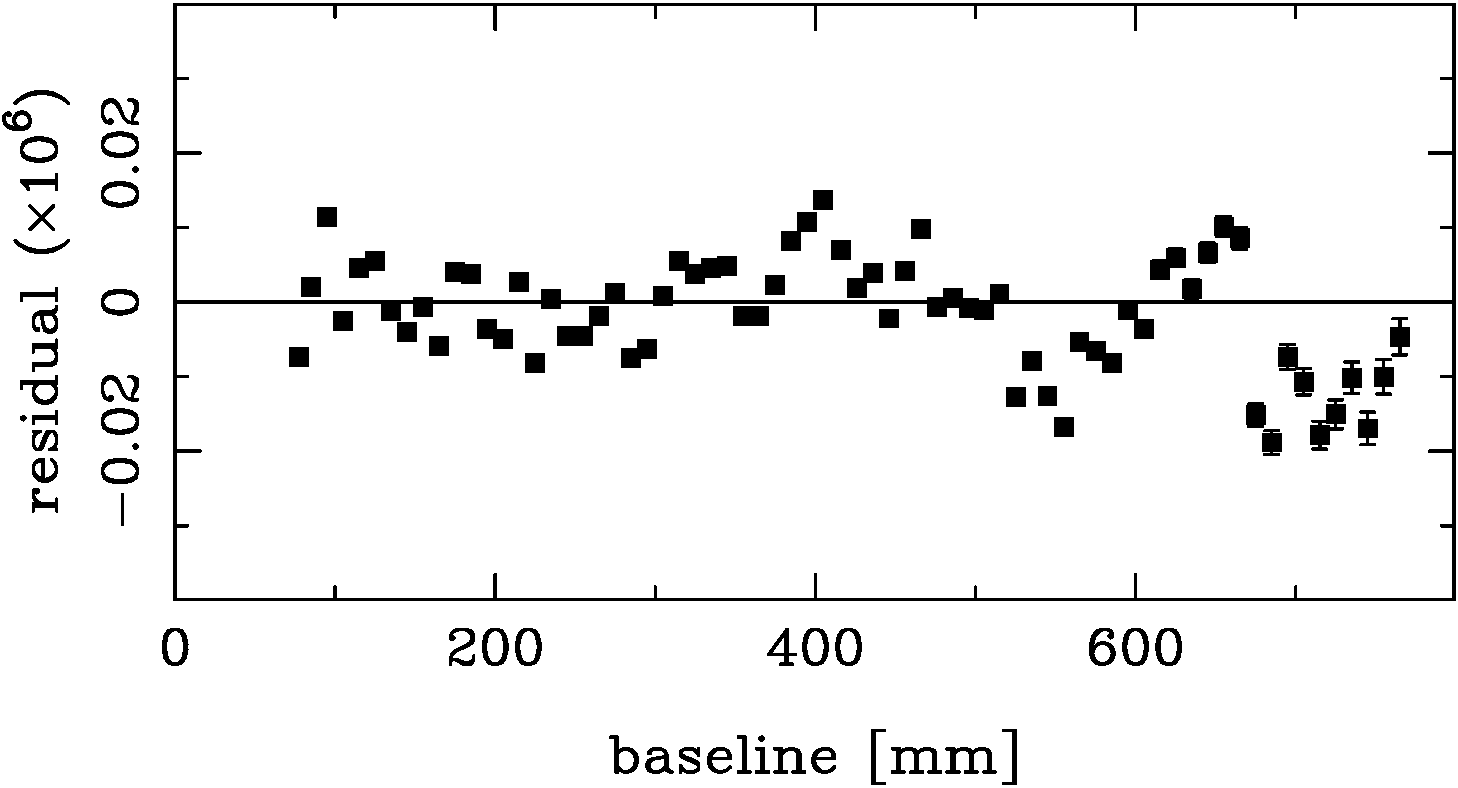}
\caption{\emph{Top panel:} Second-order coherence function $|\gamma(b)^2|$ as function of baseline, obtained when observing the SLD at a distance of 45\,cm. Error bars are smaller than the data points. The continuous red line indicates the best-fit Gaussian profile with a standard deviation of 295\,mm and amplitude $9.8\times10^{-7}$, offset by a constant value of $5.8\times10^{-8}$. \emph{Bottom panel:} Difference between the data and the best-fit Gaussian model presented in the top panel. Error bars are smaller than the data points in most cases. The residual is small in absolute terms ($<$2\% of the amplitude everywhere) but statistically significant.  \label{fig:result-base}}
\end{center}
\end{figure}

Figure~\ref{fig:result-base} presents our observed $|\gamma(b)^2|$. The data follow very closely a Gaussian profile. This was to be expected because the SLD beam profile was Gaussian and the coherence function profile corresponds to the squared Fourier transform of the beam profile. When modelling $|\gamma(b)^2|$ with a Gaussian function like
\begin{equation}
|\gamma(b)^2| = a\times\exp\left[-\frac{b^2}{2b_0^2}\right] + o
\label{eq:gausscorr}
\end{equation}
we find best-fit values $a = 9.8\times10^{-7}$, $b_0 = 295$\,mm, and $o = 5.8\times10^{-8}$. The offset $o$ approximately corresponds to the white noise bias (cf. Section~\ref{sec:zerobase}) scaled by the division by $2 \langle N_{\rm ph} \rangle G(\phi)$ (only approximately because the bias level depends itself on the light intensity). We note the presence of a small ($<$2\% of the amplitude everywhere) systematic deviation from an ideal Gaussian profile which becomes especially prominent for $b \gtrsim 500$\,mm; this is not surprising given that the SLD beam shows higher-order aberrations, giving it a crescent shape. The $|\gamma(b)^2|$ curve shows a small systematic scatter -- with an rms of $7.7\times10^{-9}$, or about 0.8\% of the amplitude $a$ -- which most likely arises from residual imperfections in our experimental setup, resulting in reflections, temperature variations, and fluctuations in the SLD luminosity.

Using our value for $b_0$, we can measure the angular size of the SLD from the relevant Fourier transform relations. The Fourier transform of a Gaussian function with standard deviation $\sigma$ is another Gaussian function with standard deviation $\sigma' = 1/\sigma$. Distances in $uv$ space are given by the baseline length in units of wavelength, $b/\lambda$. Therefore, in our case $\sigma=b_0/\lambda$ and thus $\sigma'=\theta=\lambda/b_0$, with $\theta$ being the angular half-width of the Gaussian profile of the emitter. For a (central) wavelength $\lambda=785$\,nm, we find $\theta \approx 0.55''$. At a distance of 45\,cm, this translates into a physical (Gaussian) half-width of the emitter of about $1.2\,\mu$m -- in agreement with the estimate provided by the manufacturer of an ``effective width of about $2\,\mu$m''.

\section{Discussion \label{sec:discussion}}

We developed and constructed an optical intensity interferometer that simulates an actual astronomical interferometer. To the best of our knowledge, our interferometer is the only one in the world using avalanche photodiodes in continuous (linear) photon detection mode. Using the zero-baseline configuration, we unambiguously demonstrated the ability of our setup to detect the Hanbury Brown--Twiss effect within few minutes of integration time. Our result implies that linear-mode APDs are indeed usable for intensity interferometry. We also noted a small but statistically significant, on levels of $\sim100\sigma$ and $\sim10\sigma$ for the dark noise and white noise measurements, respectively, bias toward positive correlations. The fact that the white noise bias is smaller than the dark noise bias by a factor of about 10 can be understood from Equation~(\ref{eq:correl}): an approximately constant bias in the cross correlation (the numerator) originating somewhere in the signal processing chain is increasingly diluted with increasing irradiation of uncorrelated light (which enters into the denominator). This bias is irrelevant for our experiment but requires attention once observations are performed at low light levels -- as in most astronomical observations.

Our measurements of the angular and physical sizes of the light-emitting area of the superluminescent diode demonstrate the ability of our interferometer to measure the angular size of (sufficiently bright) astronomical sources; an obvious application is the direct measurement of the angular diameters of stars. (Whereas the SLD emits a Gaussian beam, stars can be approximated as filled disks, meaning their correlation profiles are Airy functions.)

Throughout this article, we assume that the signal-to-noise ratios we observe are indeed given by Equation~(\ref{eq_snii}), i.e., given by Poisson noise. An additional limit is imposed by the number of correlated photons actually emitted by the light source (photon degeneracy limit). This limit restricts the choice of light sources and implies that astronomical targets must have thermodynamic temperatures on the order of thousands of Kelvin to be observable by intensity interferometers \citep{hanbury1958iii}. The effective temperature of our SLD follows from the Stefan--Boltzmann law applied to an emitter with an area of about $\rm (2.4~\mu m)^2$ and a radiative power of 8~mW, resulting in $T_{\rm eff} \approx 12$,500~K. For this temperature, an observing wavelength of 785~nm, a detector quantum efficiency of 83\%, an instrumental efficiency (discussed below) of 9\%, and an observing time of 10 min, Equations (28) and (29) in \citet{trippe2014} provide a photon degeneracy limited signal-to-noise ratio of about 3900 -- well in excess of the Poissonian S/N which we measured to be about 2700. Accordingly, the photon degeneracy limit does not affect our measurements.

In both the zero-baseline configuration (Section~\ref{sec:zerobase}) and the variable-baseline configuration (Section~\ref{sec:corrprofile}), we observe values for $|\gamma(0,0)|^2$ on the order of $10^{-6}$. Theoretically, we expect to find a value which is roughly given by the ratio of the coherence time of the SLD and the coherence time imposed by our electronic bandwidth (which is 100~MHz). The coherence time of the SLD was measured to be $\tau_{\rm SLD} \approx 4\times10^{-14}$\,s by the manufacturer. The coherence time of our instrument is roughly given by the inverse of the electronic bandwidth, i.e., $10^{-8}$\,s. The ratio of the two coherence times is on the order of $10^{-6}$. Accordingly, theoretical expectation and observation are in agreement.

Equation~(\ref{eq_snii}) allows us to estimate the product of spectral photon flux and instrumental efficiency, $\eta n_{\nu}$, from the observed statistical signal-to-noise ratio. To take into account that the light from the SLD is linearly polarized, we multiply the right-hand side of Equation~(\ref{eq_snii}) with a factor $(1+m_{\rm L}) = 1.85$. For $\alpha=83$\% at $\lambda=785$\,nm (Figure~\ref{fig:apddata}), $A = \rm\pi(3~mm)^2/4 = 7.1~mm^2$, $t = 10$\,min, $\Gamma^2 = 1$ for the zero-baseline configuration, $\Delta f = 100$\,MHz, and $(S/N)_s \approx 2700$ (Section~\ref{sec:zerobase}), one obtains $\eta n_{\nu} \approx\rm 1400\,s^{-1}\,m^{-2}\,Hz^{-1}$. Each APD receives about $3.8\times10^{12}$ photons per second into a narrow band ranging from 750~nm to 820~nm (Section~\ref{sec:zerobase}). This translates into a spectral photon flux onto each detector of approximately $\rm 1.6\times10^4\,s^{-1}\,m^{-2}\,Hz^{-1}$. Comparing this number to the value observed for $\eta n_{\nu}$, we find the instrumental efficiency of our table-top interferometer to be about 9\%.

The use of linear-mode APDs in astronomical interferometers is constrained by the detector noise. The APD modules we used in our experiment have a noise equivalent power (NEP) of $8\times10^{-13}\rm\,W/\sqrt{Hz}$ at $\lambda=800$\,nm, corresponding to a bandwidth-integrated power $P=8$\,nW for an electronic bandwidth of 100\,MHz. For an astronomical observation, the detector-noise limited signal-to-noise ratio can be expressed as
\begin{equation}
\label{eq:sn-nep}    
\left(\frac{S}{N}\right)_{\rm NEP} = \frac{\eta\,A\,n_{\nu}\,\Delta\nu\,\sqrt{2 t}}{P} \, h\nu
\end{equation}
where the parameters in the numerator are the ones introduced in Equation~(\ref{eq_snii}) and $h$ is Planck's constant. Assuming $P=8$\,nW, $\eta=0.2$ (as for the Narrabri interferometer), and an optical bandwidth of $\Delta\lambda=100$\,nm centered at $\lambda=800$\,nm, it follows from Equation~(\ref{eq:sn-nep}) that observations of a zeroth-magnitude star ($n_{\nu}=10^{-4}\rm\,s^{-1}\,m^{-2}\,Hz^{-1}$) require a collecting area of about 2\,m$^2$ to achieve a signal-to-noise ratio of 5 within one hour of observations. For the same parameter values, the photon-statistics limited signal-to-noise ratio is $(S/N)_{\rm s} \approx 14$ (for $\alpha=0.8$; Equation~\ref{eq_snii}); this comparison shows that it will be essential to employ low-noise APDs in astronomical interferometers. In our Hamamatsu Photonics APD modules, the noise is contributed by (1) the dark current, which scales with temperature $T$ like $1.08^T$; (2) thermal (Johnson-Nyquist) noise, which is proportional to $\sqrt{T}$; and (3) the noise factor of the internal amplifiers, which is constant. Thermal noise and amplifier noise factor make the strongest contribution to the noise in our APD modules (S. Yazaki / Hamamatsu Photonics, priv. commun.). We verified this by measuring the noise levels at 293~K and 253~K; the drop in temperature by 40~K improved the noise levels by $<$10\%, consistent with a $\sqrt{T}$ scaling. Actual science-grade detector modules should employ low-noise circuitry which leaves the dark current as the dominant component. Due to its exponential scaling with temperature, already moderate cooling can suppress the noise level effectively (e.g., cooling by 40~K reduces the dark current by a factor 22).

We note that the diameters of the active surfaces of APDs are smaller than those of photomultiplier tubes by about one order of magnitude. This places certain constraints on the surface quality of the optical light collectors of an astronomical intensity interferometer because the collector must focus the light onto the small detector surface \citep[see also Section 5.3 of][]{trippe2014}. For a detector diameter of 3~mm, an aperture of 10~m, and a focal ratio of unity, the instrumental point spread function must be smaller than about one arcminute. This is a value easily achieved by contemporary radio dishes, meaning that an astronomical intensity interferometer will be able to operate by using light collectors with the surface quality of radio telescopes.

In combination with ongoing efforts by other groups (outlined in the introduction), our results further strengthen the case for a revival of optical intensity interferometry in astronomy: with a small group and a small budget, we were able to construct a complete optical interferometer that can, at least in principle, be used for astronomical observations if connected to sufficiently large light collectors. This is in stark contrast to the massive technical efforts required for optical amplitude interferometry. Specifically, our successful proof-of-concept of an intensity interferometer using linear-mode APDs open a window of opportunity for multi-channel intensity interferometry (discussed in detail in \citealt{trippe2014}) using spectrally dispersed light; this technology will make it possible to exceed the efficiency (in terms of signal-to-noise limitations) of the Narrabri interferometer by orders of magnitude. Depending on its size, a future multi-channel intensity interferometer will be able to address science cases like the angular diameters of stars, measurements of the size of white dwarfs, matter flows in interacting binaries, starspots, and the accretion disks around supermassive black holes in active galactic nuclei.

\section*{Acknowledgements}

This project has been supported by the Korea Astronomy and Space Science Institute via grant no. 3348--20170006.

\section*{Data Availability}

The data underlying this article will be shared upon reasonable request to the corresponding author.



\bibliographystyle{mnras}
\bibliography{bibliography} 





\bsp	
\label{lastpage}
\end{document}